%% file: chi2020.tex
\documentclass{sigchi}

\toappear{\scriptsize Permission to make digital or hard copies of all or part of this work for personal or classroom use is granted without fee provided that copies are not made or distributed for profit or commercial advantage and that copies bear this notice and the full citation on the first page. Copyrights for components of this work owned by others than ACM must be honored. Abstracting with credit is permitted. To copy otherwise, or republish, to post on servers or to redistribute to lists, requires prior specific permission and/or a fee. Request permissions from permissions@acm.org. \\
{\emph{CHI '20, April 25--30, 2020, Honolulu, HI, USA.} } \\
Copyright is held by the owner/author(s). Publication rights licensed to ACM. \\
ACM ISBN 978-1-4503-6708-0/20/04\ ...\$15.00.\\
http://dx.doi.org/10.1145/3313831.3376825}

\usepackage{balance}       
\usepackage{graphics}      
\usepackage[T1]{fontenc}   
\usepackage{txfonts}
\usepackage{mathptmx}
\usepackage[pdflang={en-US},pdftex]{hyperref}
\usepackage{color,colortbl}
\usepackage{booktabs}
\usepackage{textcomp}
\usepackage{enumitem}
\usepackage{array}
\usepackage{makecell}
\usepackage{multirow}
\usepackage{tabularx}
\usepackage{xspace}
\usepackage{subcaption}
\usepackage{efbox,graphicx}
\efboxsetup{linecolor=black,linewidth=1pt}
\usepackage[export]{adjustbox}
\usepackage{microtype}        
\usepackage{ccicons}          

\usepackage{todonotes}
\usepackage{xcolor}
\definecolor{LightCyan}{rgb}{0.88,1,1}
\definecolor{LightAmber}{rgb}{0.98,0.89,0.74}
\definecolor{pink}{rgb}{0.9,0,0.9}
\definecolor{gray}{rgb}{0.95,0.95,0.85}
\definecolor{mauve}{rgb}{0.1,0.7,0.2}
\definecolor{bg}{rgb}{0.95,0.95,0.9}
\definecolor{dkgreen}{rgb}{0,0.6,0}
\definecolor{gray2}{rgb}{0.95,0.95,0.95}
\definecolor{gray3}{rgb}{0.20,0.20,0.20}

\newcolumntype{L}[1]{>{\raggedright\let\newline\\\arraybackslash}p{#1}}

\newcommand{\dronerescue}{DroneResponse\xspace}
\def\plaintitle{The Next Generation of Human-Drone Partnerships: \\Co-Designing an Emergency Response System}

\def\emptyauthor{}
\def\plainkeywords{Situational Awareness, Human-CPS interactions, Unmanned Aerial Vehicles, Emergency Response, Participatory design}

\makeatletter
\def\url@leostyle{%
  \@ifundefined{selectfont}{
    \def\UrlFont{\sf}
  }{
    \def\UrlFont{\small\bf\ttfamily}
  }}
\makeatother
\def\pprw{8.5in}
\def\pprh{11in}

\definecolor{linkColor}{RGB}{6,125,233}
\hypersetup{%
  pdftitle={\plaintitle},
  pdfauthor={\emptyauthor},
  pdfkeywords={\plainkeywords},
  pdfdisplaydoctitle=true, 
  bookmarksnumbered,
  pdfstartview={FitH},
  colorlinks,
  citecolor=black,
  filecolor=black,
  linkcolor=black,
  urlcolor=linkColor,
  breaklinks=true,
  hypertexnames=false
}

\begin{document}

\title{\plaintitle}

\numberofauthors{2}
\author{%
  \alignauthor{Ankit Agrawal, Sophia Abraham \\Benjamin Burger, Chichi Christine \\Luke Fraser, John Hoeksema, Sara Hwang, Elizabeth Travnik, Shreya Kumar \\Walter Scheirer, Jane Cleland-Huang\\
    \affaddr{Computer Science and Engineering, \\University of Notre Dame, USA}\\
    \email{\{aagrawa2,sabraha,skumar\}@nd.edu\\\{wscheirer,JaneHuang\}@nd.edu}}\\
     \alignauthor{Michael Vierhauser$^2$, Ryan Bauer$^3$, Steve Cox$^3$\\
    \affaddr{$^2$Business Informatics -- Software Engineering \\Johannes Keppler University, Linz, Austria}\\
    \affaddr{$^3$South Bend Fire Department\\South Bend, IN, USA}\\
    \email{ michael.vierhauser@jku.at, rbauer@southbendin.gov}}\\
}

\maketitle

\begin{abstract}
The use of semi-autonomous Unmanned Aerial Vehicles (UAV) to support emergency response scenarios, such as fire surveillance and search and rescue, offers the potential for huge societal benefits.  However, designing an effective solution in this complex domain represents a ``wicked design'' problem, requiring a careful balance between trade-offs associated with drone autonomy versus human control, mission functionality versus safety, and the diverse needs of different stakeholders. This paper focuses on designing for situational awareness (SA) using a scenario-driven, participatory design process. We developed SA cards describing six common design-problems, known as SA demons, and three new demons of importance to our domain. We then used these SA cards to equip domain experts with SA knowledge so that they could more fully engage in the design process. We designed a potentially reusable solution for achieving SA in multi-stakeholder, multi-UAV, emergency response applications. 
\end{abstract}





\input{pages/sec_introduction.tex}
\input{pages/sec_relatedWork.tex}

\input{pages/sec_keyflaws.tex}
\input{pages/sec_case.tex}

\input{pages/sec_SA_scenarios.tex}

\input{pages/sec_designsolutions.tex}

\input{pages/sec_pdResults.tex}

\input{pages/sec_threats.tex}
\section{Threats to Validity}
There are several threats to the validity of our study. First, we opted to capture field notes instead of recording the design sessions as we did not want any of the participants to feel inhibited. To minimize the potential for lost information we had two note-takers at each design session (see supplementary material) and used information from each set of notes to identify design contributions. Second, while we expect our results to generalize to a broader set of socio-technical CPS, they are derived from a limited number of design sessions for a single system in a specific domain.  Finally, while we have tested \dronerescue with physical UAVs, the design sessions were conducted using high-fidelity simulated UAVs. We decided not to conduct physical field tests with firefighters using the current prototype for two primary reasons related to safety and human trust in the system. A prototype, by design, is developed quickly in response to user feedback, and has not yet undergone the rigorous testing required to support field-tests.
Evaluating the design in a simulated environment 
has a greater impact on certain SA demons than others; for example, it is hard to fully evaluate the UI with respect to errant mental models and out-of-the-loop syndrome without conducting field tests with physical UAVs.   However, we are currently developing a robust, safe, and reliable version of the system, based on the design solutions that emerged from this project.

\section{Conclusion}
This paper has described our participatory design process in which we engaged domain experts in identifying and analyzing design tensions, and designing a UI to support effective SA in a socio-technical CPS where humans and UAVs partner together to support emergency response. The design produced by this process reflected the domain knowledge and vision of the firefighters as well as ideas for human-drone partnerships in collaborative mission-centric environments. We documented key design observations that emerged from the participatory design process, producing a solution that could potentially be reused across other multi-user UAV applications, or even in other applications in closely related domains \cite{DBLP:journals/ijmms/Richards00,gautam2007}.

The participatory design process not only provided a solid foundation for the design of the system, but established a shared vision for moving the project forward. In the next phase of our multi-year project, we will extend the UI to enable human operators to communicate mission goals to UAVs. 
We plan to conduct field tests with the South Bend Fire Department and to further evaluate the UI design's support for situational awareness using physical UAVs in the Spring of 2020. We have established a review board in conjunction with the South Bend firefighters and the City of South Bend, whereby we will design and develop \dronerescue through a series of simulations and field-tests (on our campus and in the downtown area of South Bend), and the review board will evaluate when \dronerescue can be safely and beneficially deployed in a live emergency mission. 


\section*{Acknowledgments}
The work in this paper is partially funded by NSF grants CNS-1931962, SHF-1741781, and with summer REU support for six of the co-authors from SHF-1647342. We also thank Kristina Davis for graphics support, and Ron Metoyer and Margaret Burnett for providing feedback.
\balance{}

\bibliographystyle{SIGCHI-Reference-Format}
\bibliography{chi2020}

\end{document}

%% file: pages/sec_introduction.tex
\section{Introduction}
\label{sec:introduction}
Recently, many innovative solutions have been imagined, developed, and deployed with the goal of enabling humans and machines to engage collaboratively in real-world tasks. Such systems incorporate aspects of both Cyber-Physical Systems (CPS)~\cite{cps-survey}, and Socio-Technical Systems (STS)~\cite{emery-SST, doi:10.1080/14639220701635470} and are characterized by close co-operation between multiple humans and machines. They can be referred to as \emph{socio-technical CPS}. One example is a system that deploys small Unmanned Aerial Vehicles (UAVs) alongside human first responders ~\cite{DBLP:conf/icse/Cleland-HuangVB18} to quickly create a 3D heat map of a burning building, detect structurally unsound hotspots, or to generate thermal imagery of smoke-filled rooms to search for people trapped inside the building~ \cite{khan:hal-02128386}.  In such scenarios, a certain degree of UAV autonomy is essential as it frees up human responders to focus on mission planning and decision-making tasks without dealing with low-level details of controlling individual UAVs. Humans, however, need to collaborate closely with the UAVs,  for example, examining video streams to evaluate potential victim sightings or setting high-level mission goals.

Designing this type of system represents a `wicked' design problem \cite{Webber73,COYNE20055}, characterized by Rittel and Webber as a unique problem with no well-formulated solution, multiple stakeholders with conflicting needs, no definitive test of a solution's validity, and little opportunity to learn by trial and error \cite{Webber73}. Addressing such problems requires designers to acquire a deep understanding of stakeholders' needs, identify conflicting goals, understand subsequent tradeoffs, and ultimately to design a solution that balances these goals and tests the solution in a real-world context \cite{Tatar:2007:DTF:1466607.1466609,DBLP:journals/tse/LamsweerdeDL98}.

Historically, many CPS failures have originated in the user interface (UI). For example, in 1988 the US Navy shot down a civilian plane with 290 people on board.  The Vincennes had entered Iranian water and mistakenly identified the Airbus as an attacking F-14 Tomcat despite the fact that the Airbus was emitting civilian signals. The mistaken identification was partially attributed to a problem in the UI which caused the operator to confuse the data of a military plane in the area with that of the civilian one~\cite{cook2012}. In fact, prior studies have shown that design problems have contributed to 60\% to 85\% of accidents in aviation and medical device domains \cite{nagel1998}, including many attributed to  ``human'' failures \cite{kohn1999}.

Some of the most common user interface design problems in socio-technical CPS are related to poor Situational Awareness (SA), defined as the ability for users to perceive, understand, and to make effective decisions \cite{endsley2017autonomous}. In this paper we therefore focus on the design challenges associated with designing effective situational awareness into our ``\dronerescue'' system. \dronerescue uses multiple UAVs to support emergency responders, and introduces novel design challenges due to the high degrees of coordination and situational awareness required between humans and semi-autonomous machines. Figure \ref{fig:river} depicts an early prototype for a river-rescue scenario.  Three UAVs are dispatched on the river (shown as yellow shields), and are tasked with planning and executing search routes. UAV $\#3$, uses image recognition to identify a potential victim in the water and streams video shown in  the left side panel. 

The \dronerescue project emerged from an ongoing collaboration with members of the South Bend fire department who currently use manually operated UAVs for emergency response. Given the novelty of deploying semi-autonomous UAVs for emergency response and the subsequent lack of existing, clearly defined, and reusable design solutions, we followed a participatory design approach~\cite{Bodker:2018:PDM:3183791.3152421}. This enabled us to benefit from firefighters' knowledge, develop a shared vision, and to fully engage the firefighters in the design and validation process.

This paper claims three core contributions. First, we present design solutions, co-designed by domain experts, to address the challenging design problem of deploying semi-autonomous UAVs in emergency response scenarios.  Second, we describe and evaluate the use of visual aids, in the form of SA cards, to engage domain experts in the process of designing and heuristically evaluating a UI for situational awareness.  Finally, we introduce three new SA design challenges of particular relevance in socio-technical CPS systems, where humans and machines interact in dynamic and novel ways. 

%% file: pages/sec_relatedWork.tex
\section{Related Work}
Our work focuses on participatory design for SA in a challenging socio-technical CPS domain. While numerous authors have described studies on participatory design \cite{Bodker:2018:PDM:3183791.3152421,6465218,Schuler:1993:PDP:563076}, there is little work at the intersection of participatory design and SA, especially in the CPS domain. The seminal work on SA demons by Endsley~\cite{ endsley2012} focused on user-centered design rather than participatory design; however, our goal is to engage domain experts as co-designers rather than to design the product for them. In one exception, Lukosch et al.~\cite{6465218}, explored the use of participatory design for serious games that provided a virtual world environment in which to develop SA skills. However, the virtual environment is very different from the life-or-death domain of Emergency Response.

Several researchers have reported case studies of SA in specific domains. For example, Breuer et al.~\cite{Breuer2009InteractionDF}, explored SA for operators in the control center of a tsunami early warning system. Similar to our work, they adopted Endsley's principles of design for SA and then used domain experts to evaluate the UI for conformance to SA guidelines. However, their domain was centered around a single control center, did not include autonomous agents partnering with humans, and did not engage users in a participatory design process. Similarly, Onal et al., applied a Situation Awareness-Oriented Design (SAOD) process to analyze, design, and evaluate the UI for an electric mining shovel \cite{SA-Mines}. They used goal-directed task analysis to understand users' goals, followed by a UI design phase, and finally a UI evaluation step based on simple assessment rubriks; however, they did not use participatory design. Many papers have explored SA in systems where operators interact with a single robot or machine~\cite{8434344,DBLP:journals/ijrr/GombolayBHS17}. While we can draw lessons from their work, SA challenges associated with a single machine differ significantly from a multi-UAV emergency response environment.  

In addition, several authors developed communication dashboards to support situational awareness across teams~\cite{PARUSH2017154}. Such dashboards could be used to support  situational awareness between multiple people and machines in \dronerescue. 
Oliveira et al.~\cite{SAEMS}, explored and evaluated ways to visualize information in support of SA. Yet other work has described novel design techniques for achieving SA. For example, Daniello et al.~\cite{DANIELLO2017529}, presented an approach that uses adaptive goal selection to achieve SA. Finally, many researchers have explored ``artificial'' situational awareness for autonomous vehicles \cite{MCAREE20177038}. While this is of direct relevance for our project, our focus in this paper has been on SA for human operators.

%% file: pages/sec_keyflaws.tex
\begin{figure}[!t]
    \centering
    \includegraphics[width=0.98\columnwidth]{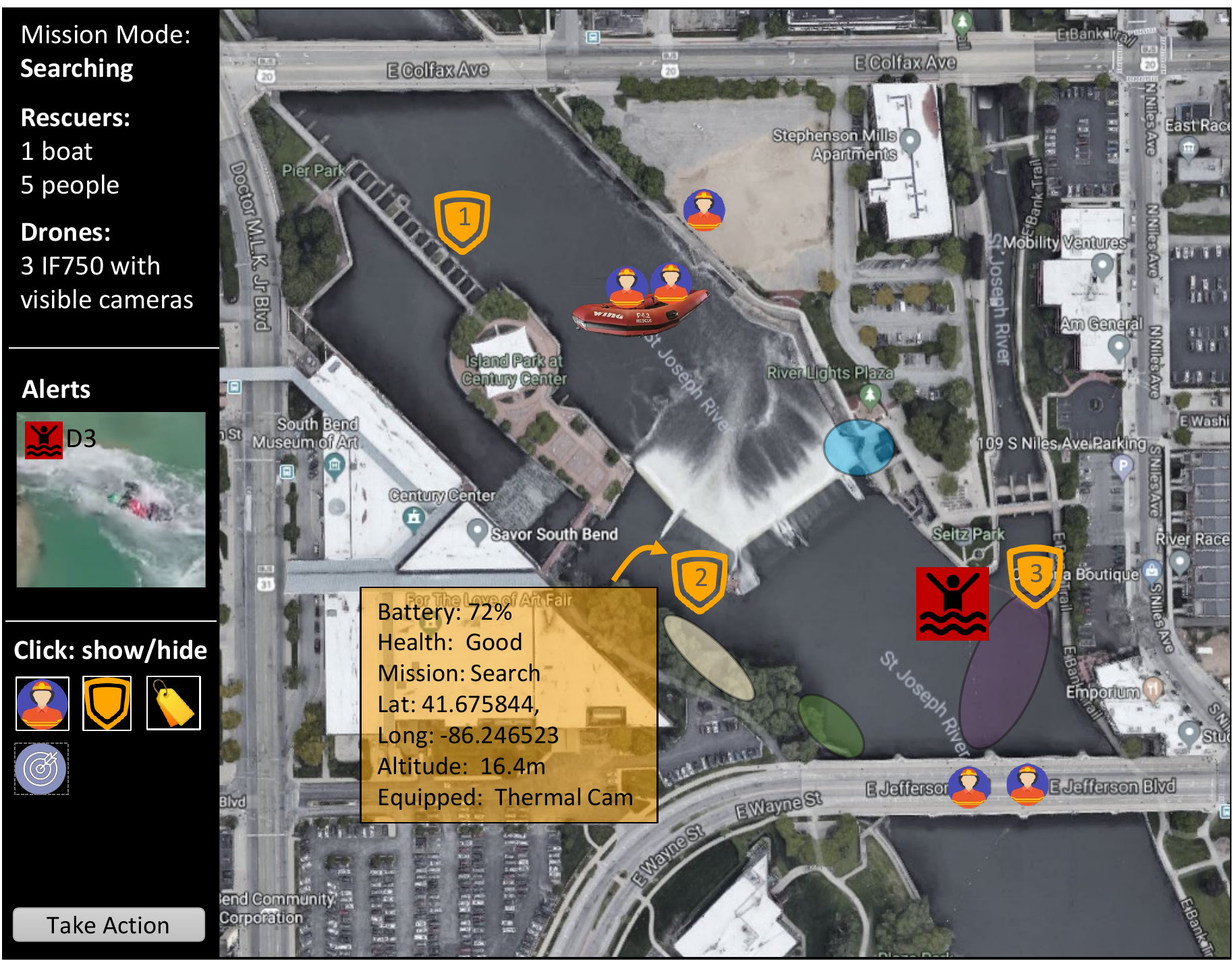}
    \caption{An Early visionary prototype  of UAV deployment for River Search-and-Rescue.}
    \label{fig:river}
    \vspace{-10pt}
 \end{figure}
 
\section{Situational Awareness}
\label{sec:barriers}
Situational Awareness is defined as the ability for users to fully perceive, understand, and make decisions in a given situation~\cite{endsley2012}. {\it Perception} (Level 1 SA) is the most basic level and involves recognizing and monitoring elements such as people, objects, and environmental factors as well as their current states. {\it Comprehension} (Level 2 SA) builds upon perception and involves the ability to develop a picture of the current situation through using pattern recognition, evaluation, and interpretation to synthesize information.  Finally, the highest level of SA, {\it Projection} (Level 3 SA), involves understanding dynamics of the environment and projection of future states. An effective UI design must support all three SA levels.
 
The basic river-rescue scenario, depicted in Figure~\ref{fig:river}, highlights several user interface challenges affecting situational awareness.  First, the three UAVs are operating semi-autonomously, and the human Incident Commander (or his/her proxy) needs full awareness of the scope of UAV permissions and the current state of the task.  This includes marking their geographical boundaries, understanding their current modes (e.g., searching, returning home), capabilities (e.g., thermal or visible-light imagery) and constraints (e.g., poor light requires thermal imagery), and empowering the designated operator(s) to issue meaningful commands.  For example, UAV $\#3$ has identified a potential victim and has automatically switched to ``track-victim'' mode. The operator is alerted to this event, inspects the video stream, confirms or refutes the victim has been found and takes appropriate actions.

\subsection{Background: Situational Awareness Demons}
There are eight common design errors that occur frequently in user interface designs which inhibit SA. These are documented as SA demons by Endsley ~\cite{endsley2012}, and are summarized below with a brief discussion of their relevance to our domain.

\subsubsection{D-1: Attentional Tunneling (AT)}
Emergency situations demand extensive awareness of the current state of the mission; however in fast-paced socio-technical CPS environments, users need to constantly process information from multiple sources. Attentional tunneling occurs when a user fixates their attention on a single informational channel for an unhealthy length of time, thereby neglecting other critical information, and failing to perform other important tasks~\cite{attentionalwickens2009,DBLP:journals/thms/RegisDRTPCT14}. For example, in the search-and-rescue scenario, attentional tunneling could occur if the design allowed the user to expand imagery streaming from one UAV in a way that entirely covered status messages from other UAVs.\vspace{-4pt}%

\subsubsection{D-2: Misplaced Salience (MS)}
In socio-technical CPS, prominent and conspicuous warnings, alerts, and alarms are often used to notify users when key events occur~\cite{prasanna09}. However, poorly placed warnings and alerts, can easily lead to misunderstandings. Our previous example of the US Naval ship shooting down a commercial plane provides a compelling example of this problem. 
\vspace{-4pt}

\subsubsection{D-3: Information Overload (IOL)}
Information overload occurs when too much information is simultaneously displayed on the screen, making it difficult for the user to process \cite{DBLP:conf/chi/KaufmanW98,McGrenere:2002:EMI:503376.503406}. For example, a user could be overloaded with multiple video streams of active UAVs, firefighter tracking information, and status messages from each UAV. To guard against information overload it is important to display only essential information and to make additional information available upon demand. Our heuristic evaluation of UIs for manually operated UAV systems (not reported in this paper) unearthed several problems with information overload, such as displaying latitude and longitude coordinates represented by a constantly changing numbers which are impossible for a human to process in real-time. \vspace{-4pt}

\subsubsection{D-4: Out-of-the-Loop Syndrome (OLS)}
Semi-autonomous systems need intermittent manual intervention to ensure safe and correct operations and must therefore be carefully and continuously monitored. However, system autonomy tends to reduce monitoring behavior of the user \cite{OutOfLoop_biondi201880}. A user is considered \textit{out-of-the-loop} if they fail to adequately monitor the system due to over-reliance on automation \cite{endsley2017autonomous}. For example, a UAV operator could fall out-of-the-loop when UAVs are autonomously planning and flying routes and the operator loses track of their current flight plans, tasks, and intents. When a user goes out-of-the-loop they will be ill-prepared to make decisions or take control in an emergency situation. Given human fallibility and the inevitability of humans losing focus, the UI must provide sufficient support to allow an operator to quickly re-enter the loop. \vspace{-4pt}

\subsubsection{D-5: Errant Mental Models (EMM)}
Over time, humans develop a mental model of the system ~\cite{mueller2011improving}. An errant mental model arises if a user's understanding of the system behavior is inconsistent with its actual behavior, creating a gap in situational awareness, and potentially causing the user to make erroneous decisions \cite{rushby2001modeling}. For example, in \dronerescue, the image recognition capabilities might be trained in good weather and daylight conditions, and an operator might incorrectly assume that it will perform at the same level of accuracy in low-visibility conditions. Maintaining a correct mental model is especially challenging in a system in which AI (artificial intelligence) is responsible for much of the underlying decision-making. \vspace{-4pt}

\subsubsection{D-6: Requisite Memory Trap (RMT)}
Humans can typically retain five to nine pieces of information in their short-term memory~\cite{DBLP:journals/tit/Miller56}; however, the information is fragile and can decay quite rapidly \cite{ATKINSON196889}. Since human short-term memory is limited, excessive reliance on it can cause intermittent loss of situational awareness. For example, it is unrealistic to expect the drone commander to remember the constantly changing state of all UAVs in the air and on the ground (e.g., their locations and remaining battery), their capabilities (e.g., onboard cameras), positions of human operators, wind conditions, and current river configuration (e.g., fallen logs) in order to make strategic search-and-rescue decisions. \vspace{-4pt}

\subsubsection{D-7: Workload, Anxiety, Fatigue, \& other Stressors (WAFOS)} WAFOS is evident when users deal with constant alarms and complex environments over time. Stratmann and Boll~\cite{demonhunt} performed a ``demon hunt'' for causes of maritime accidents and reported that 40\% of 333 studied incidents were caused by fatigue. Despite its importance, we do not initially focus on WAFOS in our study as the emergency response scenarios we are currently exploring are relatively short in duration.  \vspace{-4pt}

\subsubsection{D-8: Complexity creep (CC)} System complexity can prevent users from forming adequate internal representations of how systems function, thereby making it difficult to project future events \cite{endsley2012}. However, it is unlikely to be a near-term issue in our greenfield \dronerescue project which is in the initial phases of development and therefore contains limited features.

\subsection{New SA Demons for Socio-Technical-CPS}

In addition to Endsley's previously reported SA demons, we identified three additional demons based on our experiences of working with UAVs.  All three represent challenges that touch the physical world and therefore go beyond the boundaries of the graphical user interface and have been validated during our participatory design process.  

\subsubsection{D-9: Transition failures across Graphical \& Physical UIs (MUI)}
\begin{figure}[t]
	\centering
	\includegraphics[width=0.95\columnwidth]{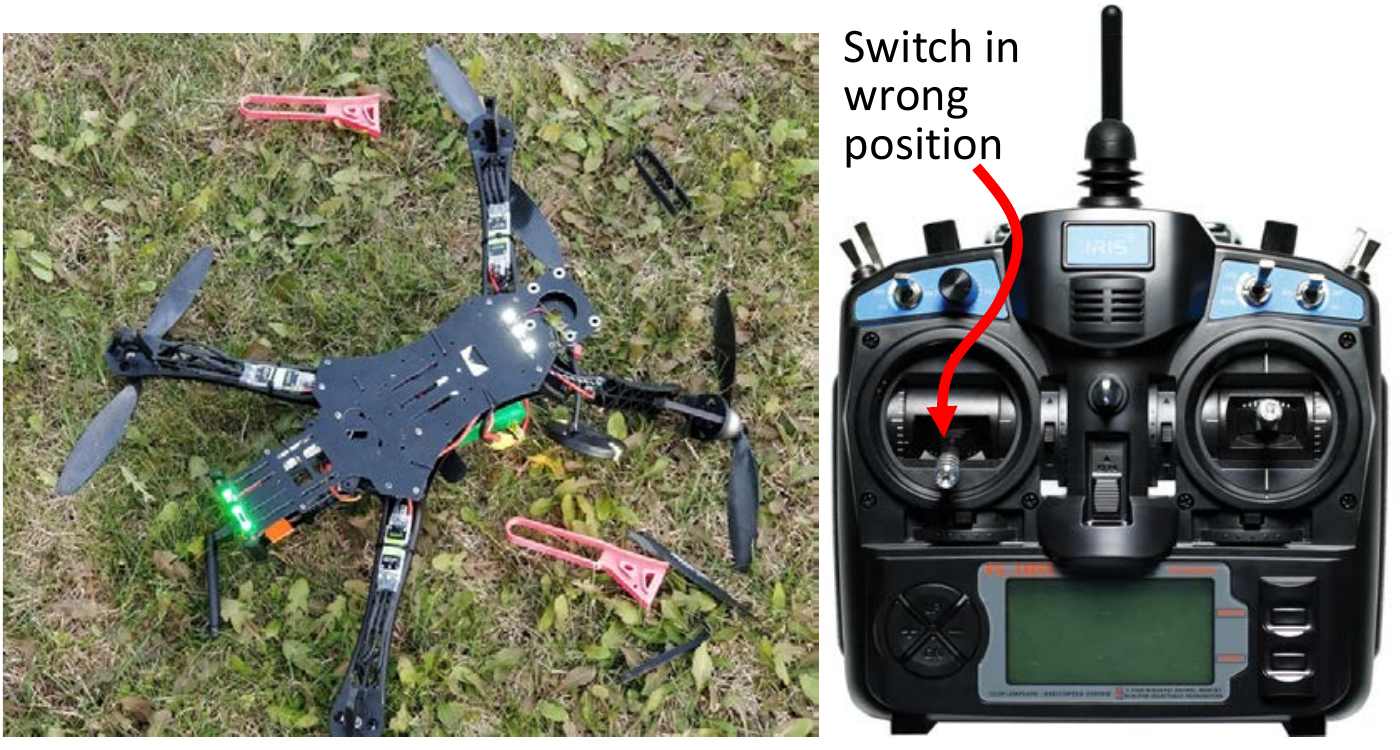}
\caption{The throttle on the hand-held device was incorrectly positioned during a transition from software-controlled flight to manually operated flight, causing the UAV to crash.}
		\label{fig:failure}
\end{figure}

We often think of situational awareness from the perspective of the graphical UI. However, a CPS often requires physical UI controls in addition to graphical ones. Misalignment of these UIs during transitions from one interface to the other can cause confusion and errors. In our prior work we described an accident that occurred when flight control was ceded from a computer to a human operator using a hand-held controller \cite{DBLP:conf/re/Cleland-HuangV18}. Prior to take-off, the operator had incorrectly positioned the throttle in the fully downward direction. This mistake was ignored during software controlled flight; however, as soon as control was ceded to the handheld controller, the UAV plummeted 30 meters to the ground and broke upon impact 
as shown in Figure \ref{fig:failure}. 
This demon is particularly insidious in our domain where UAV control is expected to pass between humans and machines.  UIs must be carefully aligned during transitions, and potential misalignments identified and mitigated through countermeasures and/or warnings.

\subsubsection{D-10: Socio-Technical CPS Communication Failure (STC)}
In a socio-technical CPS, high degrees of collaboration require clear human to human, UAV to UAV, human to UAV, and UAV to human coordination~\cite{Mccarley_humanfactors}. Communication failures across any of these pairings can isolate humans or UAVs, introduce confusion and uncertainty, reduce understanding of recent events, and force humans and/or UAVs to make independent decisions without the benefit of intended coordination mechanisms. In emergency response scenarios, communication has traditionally been based on the use of radios and hand signals, and communication failures are inevitable due to unclear spoken commands and spotty telecommunications coverage. A well-designed UI provides the opportunity to augment radio-communication with visible messaging and to provide communication logs that enable a human or UAV to quickly reconstruct situational awareness lost through a communication breakdown.\vspace{-4pt}


%

\subsubsection{D-11: Enigmatic Autonomy (EAU)}
An autonomous system can change its behavior at runtime in response to changing conditions and events \cite{88585}. Humans therefore need to understand the capabilities and permissions of the autonomous system \cite{10.3389/frobt.2018.00015}  in order to interpret its behavior. For example, the human operator must understand when and why a UAV makes an autonomous decision to return to launch (RTL), switch operating modes from \emph{search} to \emph{track-victim} mode, or change its flight plan to avoid a communication dead-spot. The UI must provide sufficient information to keep the operator aware of current autonomous permissions. 


\noindent Each of the SA demons proposed by Endsley (S1-S8) \cite{endsley2017autonomous} comes with a set of design principles that can be used to heuristically evaluate the design and to provide potential solutions for identified problems.




%% file: pages/sec_case.tex
\section{The Design Process}
\label{sec:Case}
Our design process involved six collaborative sessions including a total of 15 emergency responders and 10 researchers as summarized in Table \ref{tab:timeline}. All emergency responders were selected by the Fire Chief due to their expertise in UAV piloting and river rescue. The city of South Bend provides swift-water river rescue training to approximately 100 emergency responders across the eastern part of the USA every year. Participants involved in key design sessions, included the Fire Chief, the city's Drone Operations Coordinator (who is also a certified remote pilot (RPIC)), and six additional FAA certified RPICs.  All of these participants were trained and experienced in search-and-rescue missions. Brainstorming and ride-along activities in earlier phases of our project involved additional stakeholders with more general fire-fighting knowledge. These included a fire inspector and shift supervisors. 
\begin{table}[h]
\begin{footnotesize}
\setlength{\tabcolsep}{0.09em}
\caption{Timeline and participants of design sessions.}
\label{tab:timeline}
\begin{tabular}{|L{1.3cm}|L{4.4cm}|L{2.6cm}|}
\hline 
{\bf Date} & {\bf Participants} &{\bf Meeting Purpose}\\ \hline
04/03/19&FireChief, Drone ops Coord., Chief of Operations, 2 Researchers & Project Planning: Vision setting\\ \hline
05/23/19&FireChief, Drone ops Coord., Chief of Operations, 2 Researchers & Project Planning: Scenario creation\\ \hline
06/27/19& Fire Chief, Drone Ops Coord., 6 firefighters, 1 fire inspector, \newline 9 Researchers & Requirements Discovery: Brainstorming\\ \hline
July 2019& 6 Researchers and Firefighter shift supervisors& Ethnography: Ride-alongs\\ \hline
{\bf 07/12/19~*}&Drone ops Coord., 1 reg firefighter, 5 Researchers& Participatory design (paper prototypes)\\ \hline
{\bf 08/28/19~*}&Fire Chief, Drone Ops Coord., 3 researchers, 1 note-taker &Participatory design (exec. prototype) \\ \hline
{\bf 09/02/19~*}& Fire Chief, Drone Ops Coord., 6 Drone Ops, 4 researchers& Participatory design (exec. prototype) \\ \hline
\end{tabular}
\end{footnotesize}
\end{table}

\subsection{Requirements Discovery}
\begin{figure}[t!]
\begin{center}
\includegraphics[width=1\columnwidth]{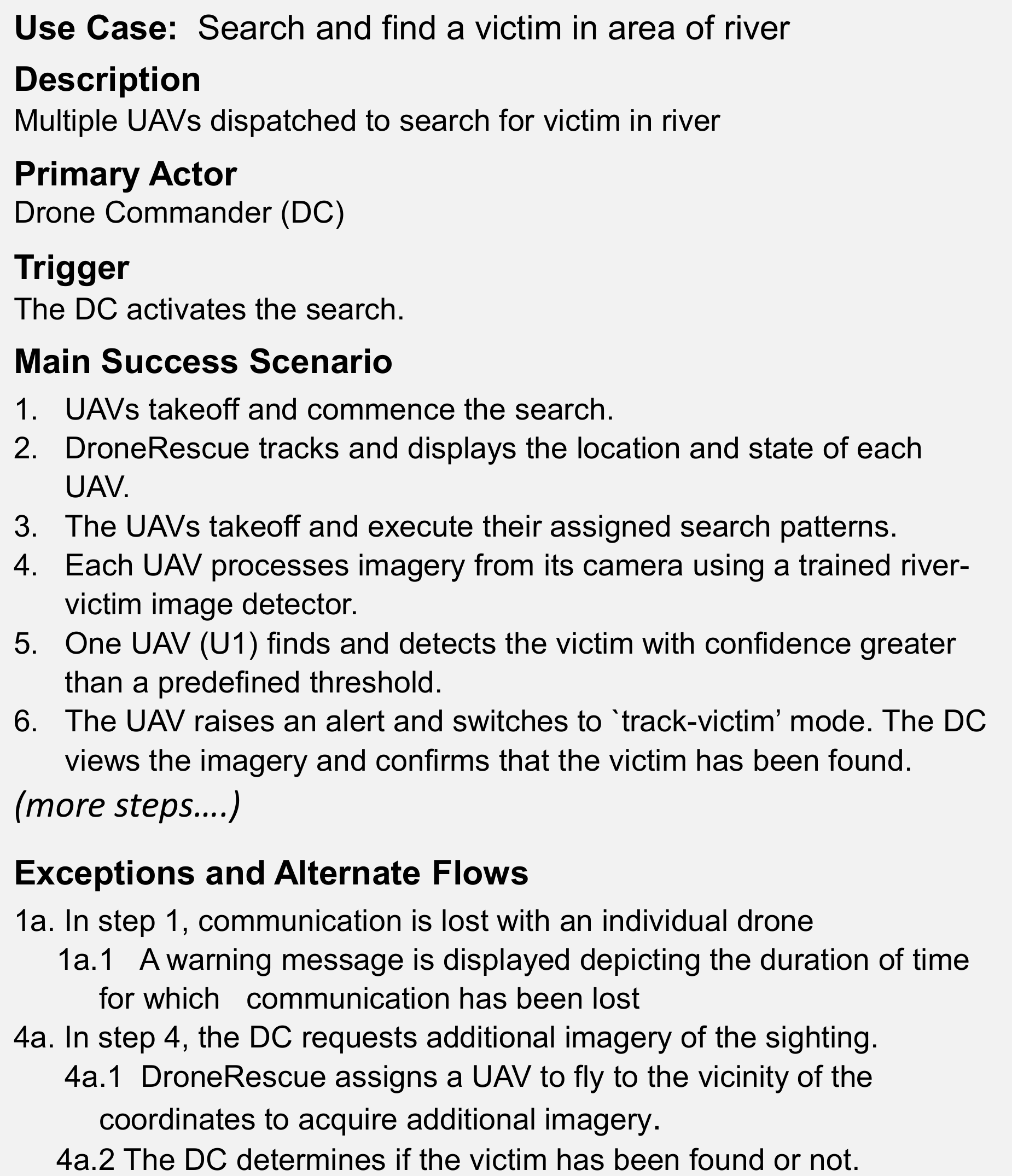}
\caption{A partial Use Case for River Rescue. The complete use case is available as supplementary material. }
\label{fig:usecase}
\end{center}
\vspace{-12pt}
\end{figure}

At the start of the project we conducted a series of meetings. The first two meetings focused on establishing a shared vision for the project, while the third meeting was used to identify and prioritize use cases of interest, including river-rescue, which is the focus of the remainder of this paper. As illustrated in Figure \ref{fig:usecase} each use case describes the main sequence of actions, as well as alternatives and exceptions~\cite{Cockburn:2000:WEU:517669} (cf. Figure~\ref{fig:usecase}). Many different scenarios can be extracted from a single use case.

\subsection{Preliminary Design and Prototype Development}
In the first design session, the firefighters show-cased their current off-the-shelf system called \emph{DroneSense}. DroneSense does not support any form of UAV autonomy, but does provide features for tracking manually flown drones manufactured by Da-Jiang Innovations (DJI). The firefighters explained which features they particularly liked or did not like, and also  missing features that would be helpful to them. This feedback, in conjunction with an initial heuristic evaluation of their system, provided insights into SA demons, such as \emph{Information Overload} and \emph{Requisite Memory Trap} that they currently experienced.  We then used a set of paper prototypes that included maps, video streams, and objects (e.g., UAVs, people, fire engines), that the firefighters manually arranged into their desired screen layout.  This activity provided general feedback about the screen layout and elements that were particularly important to the firefighters. For example, they discussed specific roles (e.g., Incident Commander vs. UAV Commander) and their informational needs, and requested new features to  track UAVs and people and to mark objects, such as fire engines or hazards, on the map. 

Based on the information collected during the first participatory design session we developed wireframes for river-search-and-rescue. We followed standard design principles based on Nielson's \cite{DBLP:conf/chi/NielsenM90} and Norman's \cite{Norman:2002:DET:2187809} principles for interaction design, as well as Endsley's design principles for avoiding SA demons \cite{endsley2012}. We created a rapid prototype as an executable Angular App, supported by existing software systems including DroneKit Python and Ardupilot to control single UAVs~\cite{ardupilot}, the Dronology platform for coordinating and monitoring multiple UAVs ~\cite{DBLP:conf/icse/Cleland-HuangVB18}, and YOLOv3, a pre-trained real-time object detection algorithm to support image recognition  ~\cite{DBLP:journals/corr/abs-1804-02767}. 

\subsection{Designing for Situational Awareness}
In the final phase, we focused on the design of \dronerescue, paying special attention to situational awareness in the context of UAV autonomy, and multi-agent communication and coordination. Less innovative aspects of the application, such as displaying UAV positions and movement on the map, were also addressed but were not a focus of our work. The remainder of this paper describes the design process and its outcomes.

%% file: pages/sec_SA_scenarios.tex
\section{Methodology: Engaging Domain Experts}
In early phases of the design process, we recognized the need for domain experts to understand concepts of situational awareness in order to engage more actively in the design process~\cite{DBLP:conf/chi/Gray16}. By exposing domain experts to SA design concepts we hoped that (1) they could provide better input as we collaboratively evaluated the emergent design, and (2) provide higher-quality design suggestions informed by their knowledge about designing for SA. We explored different solutions and adopted two techniques of (1) \emph{using visual aids} to build an understanding of situational awareness, and (2) \emph{using a scenario-based approach} to evaluate situational awareness of the preliminary design. The first technique was chosen to empower our stakeholders to recognize and discuss SA issues, while the scenarios were adopted for their proven effectiveness for engaging users in the requirements discovery and design process.\vspace{-4pt}

\subsection{Using Situational Awareness Cards to Build Knowledge}
\begin{figure}[t!]
\begin{center}
\includegraphics[width=1\columnwidth]{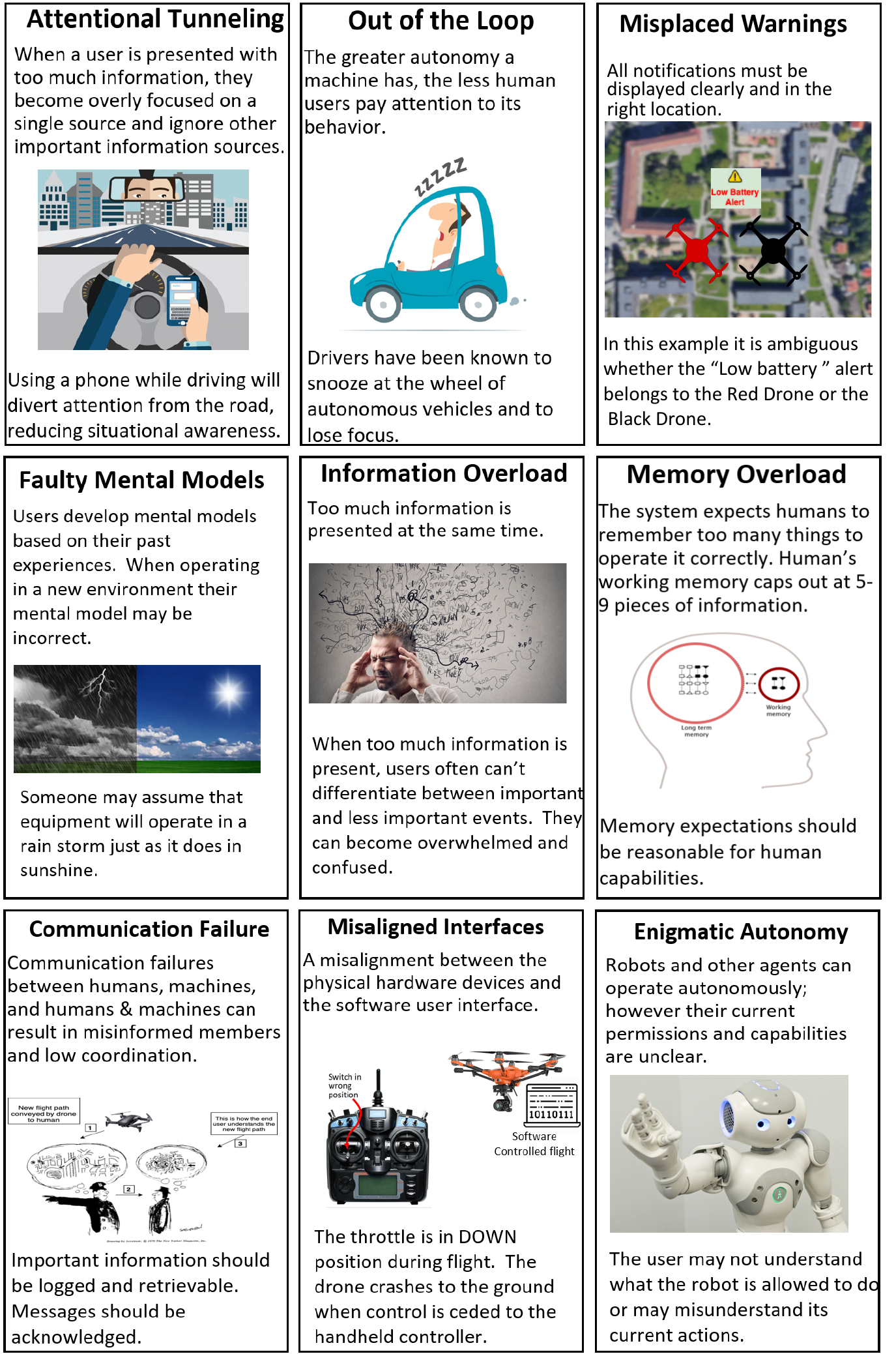}
\caption{Visual aids were developed to present concepts of Situational Awareness demons and to engage domain experts in the design process. A complete set of SA cards are available in the supplementary material.}
\label{fig:sa}
\end{center}
\end{figure}

We created a set of graphical cards describing each SA demon as depicted in Figure~\ref{fig:sa}. We simplified the names of some demons to make them more memorable, for example, changing ``Misplaced Salience'' to ``Misplaced Warnings''.  For training purposes, we briefly described the SA demon depicted in each card. After presenting each card, we reinforced the concepts by asking firefighters to think of scenarios in their own domain in which the SA demons might be particularly relevant and asked them if they had experienced any of these SA demons with their current manual system (DroneSense). Finally, we specifically reminded them that several of the SA demons are particularly relevant when UAVs exhibit autonomous behavior and that it was therefore important to rigorously validate the \dronerescue design against these demons. \vspace{-4pt}

\subsection{Scenario Driven Approach to Validation}
We derived scenarios from system use cases specified during the requirements discovery phase (e.g., Figure \ref{fig:usecase}). Use cases have multiple potential execution paths, each one representing a concrete scenario describing the system's behavior ~\cite{DBLP:journals/tsmc/Kaindl00,DBLP:conf/ACMdis/Potts95}. Developing an appropriate and sufficient set of scenarios to explore and validate the user interface design is similar to the challenge of creating a good set of user acceptance tests, where complete coverage is impossible~\cite{DBLP:conf/re/HotomskiG18,usecaseSLR}. We therefore developed a set of six guidelines for deriving scenarios from use cases to better engage domain experts in the design process. Each of the following guidelines describes an event or situation and maps the guideline back to an SA Demon.\vspace{-4pt}

\begin{itemize}[leftmargin=0cm]
    \setlength{\itemindent}{.7cm}
    \itemsep-0.15em 
    \item [{\bf G1:}] Distinct human-in-the-loop decision points (e.g., a UAV detects a potential victim and the user must either confirm the sighting, reject it, or request additional imagery from other perspectives) {\it (cf. SA Demons D-1 to D-10}). 
    \item [{\bf G2:}] Major mode changes made autonomously by a UAV (e.g., searching $\rightarrow$ track-victim, or flying $\rightarrow$ RTL (return to launch)) {\it (cf. SA Demon D-11)}.
    \item [{\bf G3:}] Situations in which the user must project future behavior in order to make a decision. {\it(i.e., Level 3 SA)}.
    \item [{\bf G4:}] Coordination between physical and graphical UIs {\it(cf. SA Demon D-9)}.
    \item [{\bf G5:}] Anticipated exception cases (e.g., communication failures with a drone, or a fly-away event when the system loses control of a UAV) {\it(cf. SA Demon D-10)}.
    \item [{\bf G6:}] Common case scenarios which could execute for significant periods of time (e.g., tracking flying UAVs) {\it(cf. SA Demon D-2)}.
\end{itemize}

Furthermore, variants of these scenarios were developed to test multiple simultaneous events {\it(cf. SA Demon D-3)}, and diverse environmental conditions (e.g., light, dark, low-visibility) {\it(cf. SA Demon D-5)}.

\subsection{Scenarios for Evaluating SA} 
The following scenarios were used to evaluate and explore the UI design.  Several of them include variants which were derived from alternate paths through the original use case. 


\noindent{\bf S1: Victim Identification:~}
(S1.1) The UAV conducts a search for a victim in the river. The UAV flies over an object, processes the image, identifies the object as a likely victim, aborts the current search, and switches to ``track-victim'' mode.  In the second related instance (S1.2), the UAV identifies an object but rejects it as a potential victim, remains in its current mode, and continues flying {\it(maps to G2, G3, G6)}.\vspace{-2pt}

\noindent{\bf S2: Lost Communication:~}
(S2.1) A UAV flies outside the expected bounds of the search area and \dronerescue loses communication with the UAV. As an exception case (S2.2), communication is lost to a UAV that remains searching within the expected search area 
{\it(maps to G5)}.\vspace{-2pt}

\noindent{\bf S3: Battery replacement:~}
(S3.1) A UAV with low battery, autonomously returns home for a battery replacement. Simultaneously, a new UAV takes-off from the home-base as a replacement. In an alternate exception case  (S3.2) a UAV with over 80\% battery remaining incorrectly returns home for a new battery {\it(maps to G2,G7)}.\vspace{-2pt}

\noindent{\bf S4: Manual Takeover:~}
Each UAV is assigned a unique color that is used to mark the physical UAV, its hand-held controller, and its UI proxy. In this scenario the system raises a malfunction alarm on the \emph{red} UAV and the operator responds by picking up the red hand-held controller; however, due to incorrect labeling (for purposes of the scenario), the controller is actually paired to the \emph{blue} UAV. As a result the blue UAV, and not the red UAV, responds to the manual flight controls. This scenario evaluates the extent to which the UI enables the user to understand what is happening despite the physical/graphical interface misalignment {\it(maps to G4,G6)}.\vspace{-2pt}

\noindent{\bf S5: Multiple Alarms:~}
Multiple alarms are raised simultaneously: (1) communication is lost with a responder's tracking device, (2) a UAV identifies a potential victim, and (3) another UAV raises a low battery alarm {\it(maps to G8)}. \vspace{-2pt}

\subsection{Using Scenarios to Co-Design for Situational Awareness}
At the start of sessions two and three, we used the SA cards to introduce SA concepts and their associated demons. We encouraged firefighters to freely contribute ideas and to share both positive and negative feedback as this would help us to  incrementally improve the design. We then systematically conducted scenario walkthroughs \cite{Lewis2006TASKCENTEREDUI}, using the following questions to trigger discussion for each scenario. 

\begin{figure*}[!t]
\centering
\begin{subfigure}[b]{.48\linewidth}
\efbox{\includegraphics[width=\linewidth, valign=t]{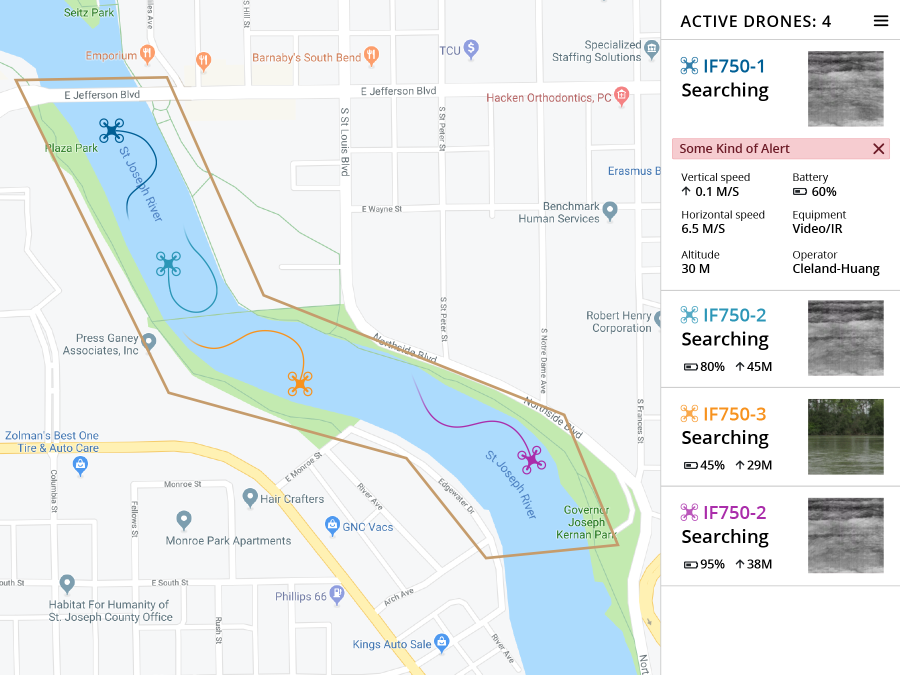}}
\caption{We developed an executable prototype from the design shown here. The prototype was used to evaluate the SA demon scenarios. Firefighters confirmed details such as color-coding of UAVs and displaying thumbnail images. They needed the color-scheme to differentiate victim-warnings from other warnings. }
\label{fig:mockup1}
\end{subfigure}
\hspace{12pt}
\begin{subfigure}[b]{.48\linewidth}
\efbox{\includegraphics[width=\linewidth,valign=t]{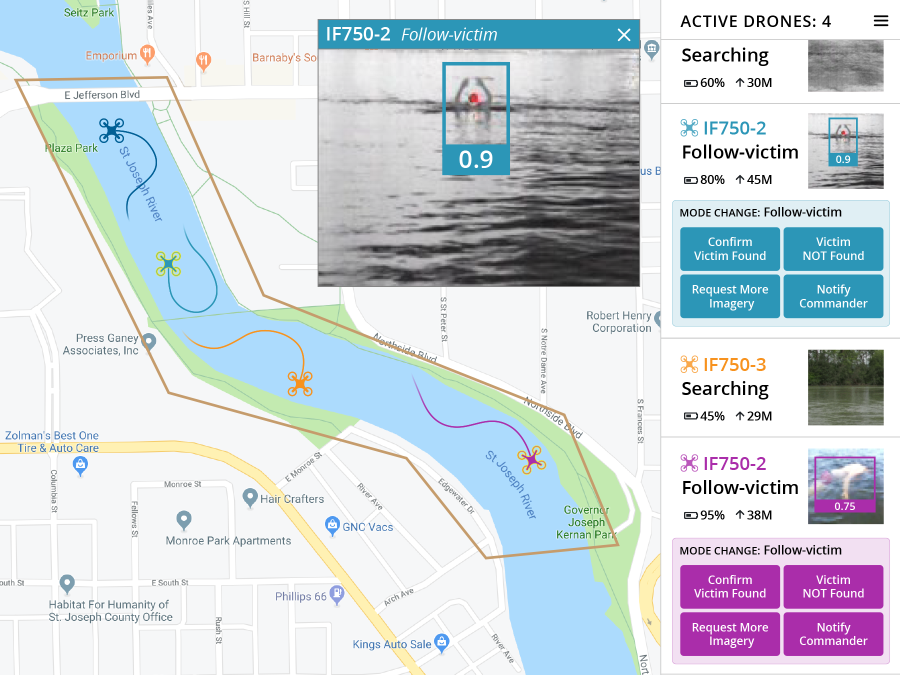}}
\caption{Firefighters requested two changes: image screens (shown here) should be replaced with a full-screen view, and action buttons should be moved from the side display to the main display. They confirmed that UAVs should autonomously switch to track-victim mode when a victim was sighted.}
\label{fig:mockup2}
\end{subfigure}
\begin{subfigure}[b]{.48\linewidth}
\vskip 0pt
\efbox{\includegraphics[width=\linewidth, valign=t]{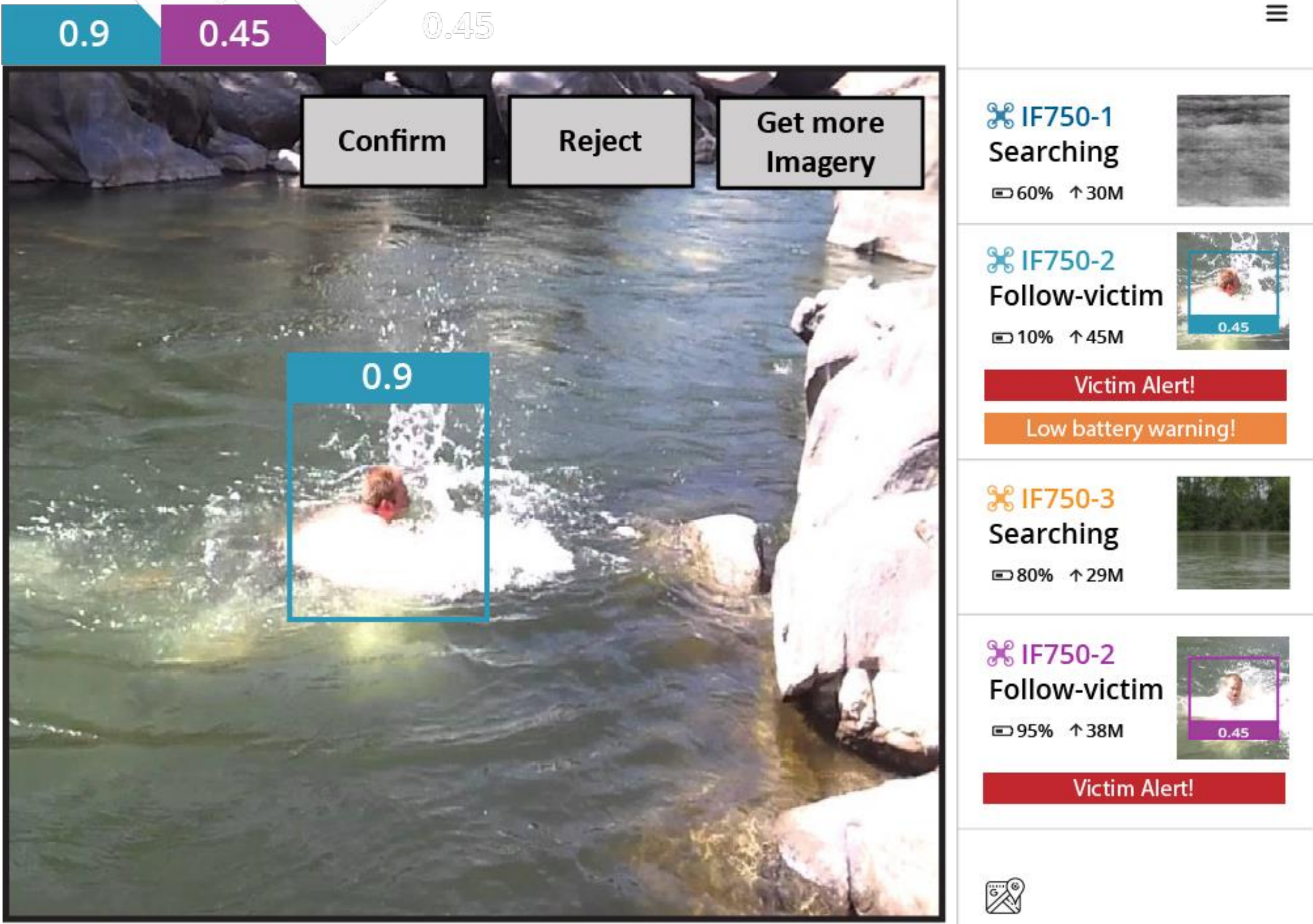}}
\caption{Video imagery was enlarged and action buttons moved to the central display. Tabs were used to depict multiple sightings and annotated with confidence scores. Bi-color warning messages were implemented. IF750-2's low battery  means it must be returned home for a battery replacement, leading to discussions about drone-autonomy vs. human-control.} 
\label{fig:mockup3}
\end{subfigure}
\hspace{12pt}
\begin{subfigure}[b]{.48\linewidth}
\vskip 0pt
\includegraphics[width=\linewidth]{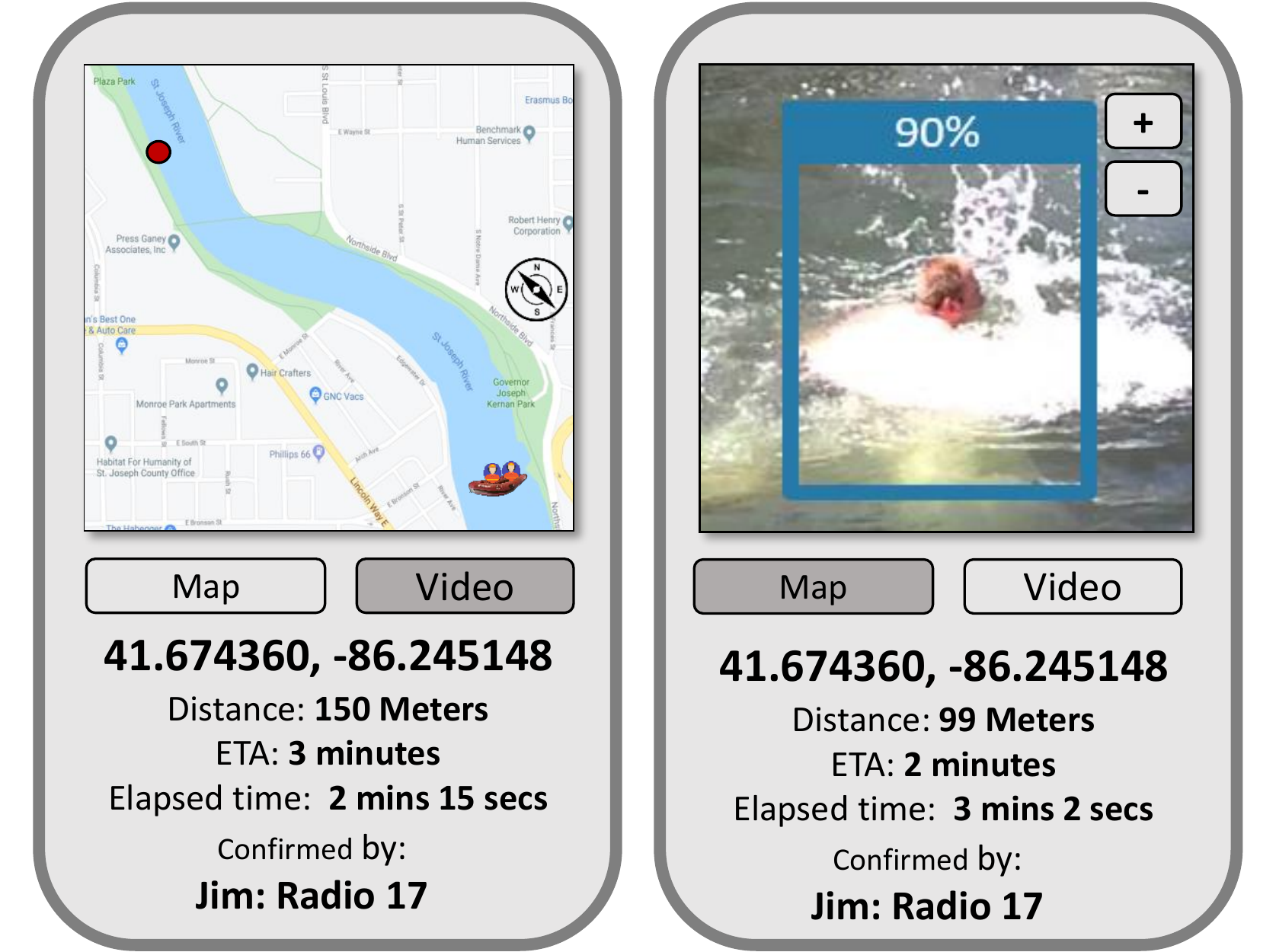}
\caption{The firefighters suggested automatically pushing a GPS location pin to rescuers on the river, whenever they confirmed a victim sighting. This suggestion illustrates the multi-role, socio-technical aspects of the domain. This prototype represents an initial mockup of the rescuers' mobile App UI showing imagery, current rescuer location, victim location, and additional information.}
\label{fig:mockup4}
\end{subfigure}
\vspace{-3pt}
\caption{The design was incrementally refined based on feedback from firefighters during co-design sessions.}
\label{fig:Design_Mockup}
\vspace{-8pt}
\end{figure*}

\begin{itemize}\itemsep-0.1em 
    \item {\bf What is happening?} This question was designed to test SA support for  \emph{perception} (i.e., Level 1 SA) and \emph{comprehension} (i.e., Level 2 SA).  For example, in scenario S2, we explored whether the users understood that communication had been lost, and later restored, to the UAV. We also asked if any important information was missing.
    \item {\bf What do you want to be able to do?} Situational awareness not only describes what is currently happening, but enables users to make decisions based on \emph{projections} of future events (i.e., Level 3 SA). For example, in scenario S1, when the UAV detected a potential victim with a certain degree of confidence, we asked the firefighters what they wanted to be able to do and how much autonomy they wanted the UAV to have.

    \item {\bf Do you see any SA demons at play?} 
    To encourage the firefighters to evaluate the design, we asked questions that probed for SA demons. For example the question ``If this scenario were enacted in rainy weather, would that change your actions (why or why not)?'' was designed to explore the Errant Mental Model demon.
\end{itemize}



%% file: pages/sec_designsolutions.tex
\section{Design Solutions for Situational Awareness}
We report the outcome of the participatory design process and address the following research question:

\noindent {\bf (RQ1)} What design solutions were adopted in \dronerescue, and which of these can potentially be reused in other UAV-based socio-technical CPS?

For each SA demon we discuss design challenges, describe important contributions made by the firefighters, and report key observations about the final design solution.  Some observations are supported by examples while others are derived from discussions which are not otherwise reported due to space constraints. Field notes from each design session are provided as supplemental materials. The design observations represent solutions that could be useful to other designers working on similar applications.  Figure \ref{fig:Design_Mockup} illustrates how one aspect of the design evolved through various stages of prototypes. 

\subsubsection{Attentional Tunneling}
Our initial design included an \emph{always visible} status panel on the right hand side of the screen which showed UAV state, thumbnail video streams, current alarms, and available operator actions. Users could request a larger video stream upon demand. However, the firefighters were adamant that when a UAV identified a victim they would exclusively focus their attention on the video stream. One firefighter stated \emph{``When we find a victim we don't care about anything else''} and proposed immediately displaying a full-screen view of the (candidate) victim.  We decided to use a tabbed view, with each tab representing a potential sighting of a victim, marked with the probability that a victim had been found. Firefighters said they would prefer to inspect image streams one-by-one, stating that \emph{``As (candidate) victims come in, we (would) prioritize the first one that comes up, clear it and go on with the next one''}. Tabs were chosen for the design because they provide a simple means for switching between similar types of views. {\bf Key Observation:}~
In this life-or-death scenario, the firefighters embraced \emph{attentional tunneling} as a necessity and did not perceive it as a demon. However, as UAVs must be continually monitored we defined the dedicated role of a technical UAV operator whose sole responsibility was to manage UAVs.

\subsubsection{Misplaced Salience}
UAV autonomy has significant impact upon the user experience in \dronerescue, especially in comparison to the current state-of-the art systems where UAVs are flown manually. This led to several design discussions focusing on the best way to display each UAV's flight information. The firefighters  proposed changes for several icons such as the route markers and flight paths. They did not initially understand the use of a blinking timer next to a UAV to depict communication loss stating {\it ``Why was it blinking? 9 secs ago? 11 secs ago?''}; however, after discussion, they agreed that blinking was a good indicator of lost communication and that they would remember what it meant in the future.  Another major difference from their previous system is that a single commander would be responsible for multiple UAVs. Scenario S5 therefore evaluated the UI's support for multiple simultaneous alarms.  The firefighters initially stated \emph{``we can prioritize the alarms ourselves''}, but then  asked \emph{``can we have two different colors? Use bright red for victim alarms so we see those first''}, and this suggestion was implemented for Design Session $\#3$.\\
\noindent{\bf Key Observation:}~
The UI needs the ability to display the UAV's recent path (i.e., its tail) as well as its planned route. Furthermore, critical victim-related alarms must be clearly differentiated from other alarms. 

\subsubsection{Information Overload}
Information overload could occur because of the need to monitor multiple  UAVs. However, we followed Endsley's design principles and as a result the firefighters did not mention information overload problems.  They did confirm several design decisions; for example, they confirmed that the common practice of displaying numerical latitude and longitude values was distracting and that the rapidly {\it``changing numbers''} were impossible to parse.  However, one firefighter said that they {\it ``might be needed to relay to dispatch''}  and  would be useful for {\it ''incidents covering wider geographical regions such as wide-area search or flood surveillance.''}  We therefore made the GPS coordinates available upon request, but retained the altitude indicator in the main UAV display.  {\bf Key Observation:}~
Avoid displaying constantly changing latitude and longitude coordinates.  Differentiate between information that must be always available in a given context (e.g., velocity and altitude), versus additional information that could be provided upon request (e.g., latitude and longitude).

\subsubsection{Out-of-the-Loop Syndrome}
An emergency response environment is noisy and fast-paced, with many distractions that could lead to a person going out-of-the-loop.  However, we did not observe this behavior during our design sessions at the fire department despite several alarms that went off during our visit.  When one UAV flew off the screen as part of a planned exception case, the  firefighters noticed that \emph{ ``it is a flyaway''} and then stopped monitoring it even though its status still appeared in the right hand panel. One firefighter stated that \emph{ ``we weren't ... snoozing, but~\ldots we did not know what was happening.''} To address this we added additional alarms, delegated the responsibility of fly-aways to the technical UAV operator, and clearly depicted current operational responsibilities in the UI. ~{\bf Key Observation:}~
The UI must enable users to recover situational awareness if they go out-of-the-loop; however, this demon is unlikely to show itself in the early phases of design and testing. We therefore defer its evaluation until the system is deployed for outdoor field tests with physical UAVs, noisy fire-trucks, and people (acting as victims) in the water.  

\subsubsection{Errant Mental Models}
The certified UAV pilots were clearly aware of the impact of environmental variables on UAV flight with comments such as  if {\it ``winds change then they (UAVs) need to land immediately and give notification''}, and stated that weather conditions needed to be clearly displayed. They also asked {\it ``if there is very little light will it (image detection) work with infrared?''} and made comments such as {\it ``you are instructing (training) the computer program to recognize''} a victim but it {\it ``will not look like this on infrared.''} The pursuing conversation showed that the firefighters understood the importance of training data matching  the current usage environment, and that they wanted to understand the impact of any mismatch. We agreed to display visual cues as reminders for different environmental conditions such as high wind, low visibility, or dark conditions, that could impact UAV behavior. ~{\bf Key Observation:}  Provide prominent visual cues, such as weather icons, for any operating conditions that are likely to impact system behavior. Make relevant information available upon request. (e.g., ``You are operating in \emph{low visibility} conditions. Expect the UAVs to fly more slowly and closer to the surface under these conditions.'')

\subsubsection{Requisite Memory Trap}
In general, we did not observe any major memory problems. This was partially because the original design displayed all available actions for every human decision point.  One minor exception occurred when a UAV automatically switched to RTL (return to launch) mode due to low battery and the firefighter tried to figure out where it was going and asked \emph{``where is home?''} To remediate this problem he proposed adding an additional icon to mark the home location of each UAV.~{\bf Key Observation:}~
Clearly identify and display any information that is needed within the current context (e.g., a UAV needs to return home, and therefore the home location is displayed on the map). Whenever operator input is needed for a critical decision (e.g., when a victim is found), display common options prominently on the screen. 

\subsubsection{Socio-Technical CPS Communication Failure}
The firefighters described the high volume of radio-chatter needed to coordinate activities during an incident, and the inevitable communication failures that occur in practice.  They saw the opportunity to \emph{``reduce radio chatter''} by creating alternate communication paths. They wanted the system to automatically  \emph{``drop a pin''} at the victim's location, and \emph{``push the pin''} to the rescuers on-board the boat whenever a firefighter confirmed a victim sighting. They also wanted the new system to further reduce radio communication by tracking positions of firefighters on the scene. ~{\bf Key Observation:}~ 
When contact is lost with either a human or a UAV, a warning must be displayed indicating the duration of lost communication and the last known position. Identify potential interfaces that could strengthen system-wide communication. 

\subsubsection{Misaligned Graphical \& Physical UIs}
Scenario S5 was specifically designed to test the alignment of graphical UIs and physical devices. In one example, the firefighters suggested the use of permanently \emph{``colored hand-held controllers''} to reduce configuration mismatches. This aligned with a previous design decision to assign each physical UAV a color code that would be stored on its permanently attached companion computer. The stakeholders also suggested providing an \emph{``ipad display on the (handheld) controller''} on each UAV to display status and to stream imagery from any onboard camera. They said this would help the remote pilots control the UAV and bring it safely back home. These issues raised during the design session reinforced the need for close pairing between physical and graphical UIs.~{\bf Key Observation:}~ 
Identify all potential pairings between graphical and physical UIs (e.g., throttle position of hand-held controller as UAV transitions from computer to manual control) and integrate consistency checks. For example, sound an alarm on the UAV Technician's UI and prohibit a UAV from taking off until its handheld controls are positioned correctly.


\subsubsection{Enigmatic Autonomy}
The firefighters raised many questions about the UAVs autonomy stating that {\it``the autonomous part is so new to us.''} However, as the project proceeded they asked increasingly informed questions that reinforced the value of participatory design. In the first walkthrough of scenario S2, a firefighter asked \emph{``How do we know that the drone is recognizing a person on the water''} or \emph{ ``holding onto a branch~\ldots as opposed to first responders standing on the river bank?''} In another scenario walkthrough, a firefighter asked \emph{``Why is it (the UAV) flying there?''}  highlighting their need to understand the UAV's permissions and capabilities in order to understand what is happening and why it is happening, and to make decisions based on their projections of future UAV behavior.\\
\noindent{\bf Key Observation:}~ 
Explanations of UAV permissions within the current environment (i.e., What is the UAV allowed to do right now?) must be provided. While not directly discussed, we also propose providing support for retrospective analysis that explains UAV actions taken during a mission  (e.g., Why did the UAV behave as it did and with what degree of confidence did it make those decisions?).



%% file: pages/sec_pdResults.tex
\section{How Participatory Design Worked in Practice}

In addition to exploring relevant design solutions, we also evaluated the effectiveness of the participatory design process itself and the extent to which the firefighters contributed constructively to the design.  We primarily focused on addressing the following research question: 

\noindent {\bf (RQ2)} To what extent did the domain experts contribute to the design  process, and is there any evidence that their exposure to SA concepts contributed to their engagement?

\begin{table}[t]
\centering
\caption{Codebook used to map design contributions to SAs.}
   \small\addtolength{\tabcolsep}{-4.5pt}
   \begin{tabular}{|l|L{2.8cm}|L{4.9cm}|}
   \hline
   \multicolumn{2}{|l|}{\bf Situational Demon}&{\bf Reference terms:}\\ \hline
   AT&Attentional Tunneling&Focus \\ \hline
   OLS&Out of the loop synd.& \textit{``what's happening''}\\ \hline
   MS&Misplaced salience&Colors,Flashing elements, Sounds, Alarms \\ \hline
   IOL&Information overload &Busy Screen, Cluttered view \\ \hline
   EMM&Errant mental model&Environmental Context \\ \hline
   RMT&Requisite memory trap &Markers and labels on maps \\ \hline
   MUI&Misaligned UI &Hardware-software-mix-up \\ \hline
   STC&Socio Technical Comm.&Drone communication, Information sharing \\ \hline
   EAU&Enigmatic Autonomy&AI,Image recognition, Autopilot, sensors \\ \hline
  
    \end{tabular}
\label{tab:codebook}
\end{table}

\subsection{Design Contributions: Analysis from Field Notes}
To answer RQ2, two researchers carefully reviewed the field notes from the three participatory design sessions reported in Table \ref{tab:timeline}, and coded each of the firefighters' suggestions according to the following categories:

\begin{itemize}\itemsep-0.35em 
    \item {\bf Novel design suggestions: } Design ideas that went beyond the existing design.   
    \item {\bf Confirmation of existing design decisions: } Confirmatory comments about features that were already implemented in the design (i.e., in either the wireframe or executable prototype). In many cases, the confirmation was supported by additional rationales.
    \item {\bf Emergent Requirements: } System level requirements and comments about problems or challenges that should be addressed, but without associated design solutions. We also differentiated between suggestions that were ultimately \emph{accepted} or \emph{rejected} as valid and feasible requirements.
    \item {\bf Other: } All other comments, clarification questions, and suggestions that did not clearly fall into one of the above categories was implicitly labeled as `other'. 
\end{itemize}

From now on we refer to these suggestions collectively as \emph{design contributions}, with the exception of those in the `other' category which are excluded from the subsequent analysis. We used deductive coding \cite{fereday2006demonstrating} to cross-reference each design contribution with its associated SA Demon(s) the code-book depicted in Table \ref{tab:codebook}. The codebook was developed iteratively through identifying key terms that served as indicators of each SA concept.  In the case of OLS we  only found one example, and therefore include the entire phrase used by the firefighters.
Second, we used an inductive coding approach \cite{thomas2006general} to identify core themes.  Two researchers individually tagged each contribution with one or more potential themes, producing 35 candidate themes, and then performed card-sorting to produce ten themes: adjacent systems, alerts, autonomy, drone operations, environment, human decisions, image recognition, rescue operations, teams \& roles, and UI design. Analyzing the encoded notes led to the following observations.

Firefighters drew on their domain knowledge in combination with SA concepts to contribute design ideas. For example, they explained that placing the buttons for responding to a potential victim sighting over the video stream would help them make faster decisions thereby indirectly referencing improved salience (displaying UI elements that support human actions within the focus area), requisite memory trap (reminding them of their immediate choices), and enigmatic autonomy (interjecting human decision making at a critical juncture of the UAV's autonomy).  They brought broad-ranging domain knowledge to the design process, thinking far outside the box of what we had previously imagined by proposing novel integration points to enhance socio-technical communication and raising the need for the image recognition algorithm to differentiate between victims and rescuers or bystanders. 

SA cards were used in Sessions two and three with the goal of increasing the level of engagement in design discussions.  All three sessions were of similar duration.  Session one (without benefit of the cards) resulted in only one design contribution with four indirect references to SA concepts, while Session two resulted in five design contributions with 19 SA references, and Session three produced six design contributions with 29 SA references.  While the number of references to SA concepts clearly increased after the cards were introduced; these, results are based on limited meetings and controlled studies would be needed to make more general claims.

\input{pages/sec_participation}

%% file: pages/sec_participation.tex
\subsection{Firefighters' Perception of Participatory Design}
We created a short questionnaire to collect feedback from the firefighters about the participatory design process ~\cite{mcnally2016children}.  We asked the firefighters to respond to two prompts: \emph{The situational awareness cards helped me to (1) evaluate the design, and (2) describe my design ideas}, using a 5-point Likert scale ranging from `Strongly Agree' (5pts) to `Strongly Disagree' (1 pt) with an additional `Not applicable' option.    

Five people strongly agreed, and two people moderately agreed that the SA cards helped them describe their design ideas, while only three strongly agreed and four moderately agreed that the SA cards helped them to evaluate the design itself. In both cases one person returned a neutral response (a different person in each case). These responses supported our observations that the firefighters proposed design solutions that addressed specific SA demons; but less frequently commented on whether the design fully mitigated the demon.

%% file: chi2020.bbl

\begin{thebibliography}{00}


\ifx \showCODEN    \undefined \def \showCODEN     #1{\unskip}     \fi
\ifx \showDOI      \undefined \def \showDOI       #1{{\tt DOI:}\penalty0{#1}\ }
  \fi
\ifx \showISBNx    \undefined \def \showISBNx     #1{\unskip}     \fi
\ifx \showISBNxiii \undefined \def \showISBNxiii  #1{\unskip}     \fi
\ifx \showISSN     \undefined \def \showISSN      #1{\unskip}     \fi
\ifx \showLCCN     \undefined \def \showLCCN      #1{\unskip}     \fi
\ifx \shownote     \undefined \def \shownote      #1{#1}          \fi
\ifx \showarticletitle \undefined \def \showarticletitle #1{#1}   \fi
\ifx \showURL      \undefined \def \showURL       #1{#1}          \fi

\bibitem{88585}
{P.~J. {Antsaklis}}, {K.~M. {Passino}}, {and} {S.~J. {Wang}}. 1991.
\newblock \showarticletitle{An introduction to autonomous control systems}.
\newblock {\em IEEE Control Systems Magazine\/} {11}, 4 (June 1991), 5--13.
\newblock
\showISSN{1066-033X}
\showDOI{%
\url{http://dx.doi.org/10.1109/37.88585}}


\bibitem{ardupilot}
{ArduPilot}. 2019.
\newblock \url{http://ardupilot.org}.   (2019).
\newblock
\newblock
\shownote{[Online; accessed 01-September-2019].}


\bibitem{ATKINSON196889}
{R.C. Atkinson} {and} {R.M. Shiffrin}. 1968.
\newblock \showarticletitle{Human Memory: A Proposed System and its Control
  Processes}.
\newblock Psychology of Learning and Motivation, Vol.~2. Academic Press, 89 --
  195.
\newblock
\showISSN{0079-7421}
\showDOI{%
\url{http://dx.doi.org/10.1016/S0079-7421(08)60422-3}}


\bibitem{OutOfLoop_biondi201880}
{Francesco~N Biondi}, {Monika Lohani}, {Rachel Hopman}, {Sydney Mills}, {Joel~M
  Cooper}, {and} {David~L Strayer}. 2018.
\newblock \showarticletitle{80 MPH and out-of-the-loop: Effects of real-world
  semi-automated driving on driver workload and arousal}. In {\em Proceedings
  of the Human Factors and Ergonomics Society Annual Meeting}, Vol.~62. SAGE
  Publications, 1878--1882.
\newblock


\bibitem{Bodker:2018:PDM:3183791.3152421}
{Susanne B{\o}dker} {and} {Morten Kyng}. 2018.
\newblock \showarticletitle{Participatory Design That Matters -- Facing the Big
  Issues}.
\newblock {\em ACM Trans. Comput.-Hum. Interact.\/} {25}, 1, Article 4 (2018),
  4:1--4:31 pages.
\newblock
\showISSN{1073-0516}
\showDOI{%
\url{http://dx.doi.org/10.1145/3152421}}


\bibitem{Breuer2009InteractionDF}
{Henning Breuer}, {Frank Heidmann}, {Mitsuji Matsumoto}, {Ulrich Raape}, {and}
  {Monika Wnuk}. 2009.
\newblock \showarticletitle{Interaction Design for Situation
  Awareness-Eyetracking and Heuristics for Control Centers}.
\newblock


\bibitem{DBLP:conf/re/Cleland-HuangV18}
{Jane Cleland{-}Huang} {and} {Michael Vierhauser}. 2018.
\newblock \showarticletitle{Discovering, Analyzing, and Managing Safety Stories
  in Agile Projects}. In {\em 26th {IEEE} International Requirements
  Engineering Conference}. 262--273.
\newblock
\showDOI{%
\url{http://dx.doi.org/10.1109/RE.2018.00034}}


\bibitem{DBLP:conf/icse/Cleland-HuangVB18}
{Jane Cleland{-}Huang}, {Michael Vierhauser}, {and} {Sean Bayley}. 2018.
\newblock \showarticletitle{Dronology: an incubator for cyber-physical systems
  research}. In {\em Proceedings of the 40th International Conference on
  Software Engineering: New Ideas and Emerging Results}. 109--112.
\newblock
\showDOI{%
\url{http://dx.doi.org/10.1145/3183399.3183408}}


\bibitem{Cockburn:2000:WEU:517669}
{Alistair Cockburn}. 2000.
\newblock {\em Writing Effective Use Cases\/} (1st ed.).
\newblock Addison-Wesley Longman Publishing Co., Inc., Boston, MA, USA.
\newblock
\showISBNx{0201702258}


\bibitem{cook2012}
{{Nancy J.} Cook}. 2007.
\newblock {\em Stories of Modern Technology Failures and Cognitive Engineering
  Successes}.
\newblock CRC Press, 2007.
\newblock


\bibitem{COYNE20055}
{Richard Coyne}. 2005.
\newblock \showarticletitle{Wicked problems revisited}.
\newblock {\em Design Studies\/} {26}, 1 (2005), 5 -- 17.
\newblock
\showISSN{0142-694X}
\showDOI{%
\url{http://dx.doi.org/https://doi.org/10.1016/j.destud.2004.06.005}}


\bibitem{DANIELLO2017529}
{Giuseppe D'Aniello}, {Vincenzo Loia}, {and} {Francesco Orciuoli}. 2017.
\newblock \showarticletitle{Adaptive Goal Selection for improving Situation
  Awareness: the Fleet Management case study}.
\newblock {\em Procedia Computer Science\/}  {109} (2017), 529 -- 536.
\newblock
\showISSN{1877-0509}
\showDOI{%
\url{http://dx.doi.org/10.1016/j.procs.2017.05.332}}
\newblock
\shownote{8th International Conference on Ambient Systems, Networks and
  Technologies, ANT-2017 and the 7th International Conference on Sustainable
  Energy Information Technology.}


\bibitem{emery-SST}
{F.E. Emery} {and} {E.L. Trist}. 1960.
\newblock \showarticletitle{Socio-technical systems}. In {\em In: Churchman,
  C.W., Verhulst, M. (Eds.), Management Science Models and Techniques}, Vol.~9.
  Pergamon, 83--97.
\newblock


\bibitem{endsley2012}
{Mica~R. Endsley}. 2011.
\newblock {\em Designing for Situation Awareness: An Approach to User-Centered
  Design, Second Edition\/} (2nd ed.).
\newblock CRC Press, Inc., Boca Raton, FL, USA.
\newblock
\showISBNx{1420063553, 9781420063554}


\bibitem{endsley2017autonomous}
{Mica~R Endsley}. 2017.
\newblock \showarticletitle{Autonomous driving systems: A preliminary
  naturalistic study of the Tesla Model S}.
\newblock {\em Journal of Cognitive Engineering and Decision Making\/} {11}, 3
  (2017), 225--238.
\newblock


\bibitem{8434344}
{K. {Fellah}} {and} {M. {Guiatni}}. 2019.
\newblock \showarticletitle{Tactile Display Design for Flight Envelope
  Protection and Situational Awareness}.
\newblock {\em IEEE Transactions on Haptics\/} {12}, 1 (Jan 2019), 87--98.
\newblock
\showISSN{1939-1412}
\showDOI{%
\url{http://dx.doi.org/10.1109/TOH.2018.2865302}}


\bibitem{fereday2006demonstrating}
{Jennifer Fereday} {and} {Eimear Muir-Cochrane}. 2006.
\newblock \showarticletitle{Demonstrating rigor using thematic analysis: A
  hybrid approach of inductive and deductive coding and theme development}.
\newblock {\em International journal of qualitative methods\/} {5}, 1 (2006),
  80--92.
\newblock


\bibitem{gautam2007}
{Naveen Gautam}, {Ratna~Babu Chinnam}, {and} {Nanua Singh}. 2007.
\newblock \showarticletitle{Design reuse framework: A perspective for lean
  development}.
\newblock {\em Int. J. Product Development Int. J. Product Development\/}  {4}
  (01 2007), 485--507.
\newblock
\showDOI{%
\url{http://dx.doi.org/10.1504/IJPD.2007.013044}}


\bibitem{DBLP:journals/ijrr/GombolayBHS17}
{Matthew~C. Gombolay}, {Anna Bair}, {Cindy Huang}, {and} {Julie~A. Shah}. 2017.
\newblock \showarticletitle{Computational design of mixed-initiative
  human-robot teaming that considers human factors: situational awareness,
  workload, and workflow preferences}.
\newblock {\em I. J. Robotics Res.\/} {36}, 5-7 (2017), 597--617.
\newblock
\showDOI{%
\url{http://dx.doi.org/10.1177/0278364916688255}}


\bibitem{DBLP:conf/chi/Gray16}
{Colin~M. Gray}. 2016.
\newblock \showarticletitle{"It's More of a Mindset Than a Method": {UX}
  Practitioners' Conception of Design Methods}. In {\em Proceedings of the 2016
  {CHI} Conference on Human Factors in Computing Systems}. 4044--4055.
\newblock
\showDOI{%
\url{http://dx.doi.org/10.1145/2858036.2858410}}


\bibitem{DBLP:conf/re/HotomskiG18}
{Sofija Hotomski} {and} {Martin Glinz}. 2018.
\newblock \showarticletitle{A Qualitative Study on using GuideGen to Keep
  Requirements and Acceptance Tests Aligned}. In {\em 2018 IEEE 26th
  International Requirements Engineering Conference (RE)}. IEEE, 29--39.
\newblock


\bibitem{DBLP:journals/tsmc/Kaindl00}
{Hermann Kaindl}. 2000.
\newblock \showarticletitle{A design process based on a model combining
  scenarios with goals and functions}.
\newblock {\em {IEEE} Trans. Systems, Man, and Cybernetics, Part {A}\/} {30}, 5
  (2000), 537--551.
\newblock
\showDOI{%
\url{http://dx.doi.org/10.1109/3468.867861}}


\bibitem{DBLP:conf/chi/KaufmanW98}
{Leah Kaufman} {and} {Brad Weed}. 1998.
\newblock \showarticletitle{Too much of a good thing?: Identifying and
  resolving bloat in the user interface}. In {\em {CHI} 98 Conference Summary
  on Human Factors in Computing Systems}. 207--208.
\newblock
\showDOI{%
\url{http://dx.doi.org/10.1145/286498.286693}}


\bibitem{khan:hal-02128386}
{Md. Nafiz~Hasan Khan} {and} {Carman Neustaedter}. 2019.
\newblock \showarticletitle{{Exploring Drones to Assist Firefighters During
  Emergencies}}. In {\em {1st International Workshop on Human-Drone
  Interaction}}. {Ecole Nationale de l'Aviation Civile [ENAC]}, Glasgow, United
  Kingdom.
\newblock
\showURL{%
\url{https://hal.archives-ouvertes.fr/hal-02128386}}


\bibitem{kohn1999}
{{L.T.} Kohn}, {{J.M.} Corrigan}, {and} {{M.s.} Donaldson}. 1999.
\newblock \showarticletitle{To err is human, Building a safety health system}.
\newblock {\em Washington, DC: National Academy Press\/} (1999).
\newblock


\bibitem{Lewis2006TASKCENTEREDUI}
{Clayton Lewis} {and} {John Rieman}. 2006.
\newblock \showarticletitle{Task-centered user interface design: A practical
  introduction}.
\newblock


\bibitem{6465218}
{H. {Lukosch}}, {T. {van Ruijven}}, {and} {A. {Verbraeck}}. 2012.
\newblock \showarticletitle{The participatory design of a simulation training
  game}. In {\em Proceedings of the 2012 Winter Simulation Conference (WSC)}.
  1--11.
\newblock
\showISSN{0891-7736}
\showDOI{%
\url{http://dx.doi.org/10.1109/WSC.2012.6465218}}


\bibitem{MCAREE20177038}
{Owen McAree}, {Jonathan~M. Aitken}, {and} {Sandor~M. Veres}. 2017.
\newblock \showarticletitle{Towards artificial situation awareness by
  autonomous vehicles}.
\newblock {\em IFAC-PapersOnLine\/} {50}, 1 (2017), 7038 -- 7043.
\newblock
\showISSN{2405-8963}
\showDOI{%
\url{http://dx.doi.org/10.1016/j.ifacol.2017.08.1349}}
\newblock
\shownote{20th IFAC World Congress.}


\bibitem{Mccarley_humanfactors}
{Jason~S. Mccarley} {and} {Christopher~D. Wickens}.
\newblock {\em Human factors concerns in UAV flight}.
\newblock {T}echnical {R}eport.
\newblock


\bibitem{McGrenere:2002:EMI:503376.503406}
{Joanna McGrenere}, {Ronald~M. Baecker}, {and} {Kellogg~S. Booth}. 2002.
\newblock \showarticletitle{An Evaluation of a Multiple Interface Design
  Solution for Bloated Software}. In {\em Proceedings of the SIGCHI Conference
  on Human Factors in Computing Systems}. ACM, 164--170.
\newblock
\showISBNx{1-58113-453-3}
\showDOI{%
\url{http://dx.doi.org/10.1145/503376.503406}}


\bibitem{mcnally2016children}
{Brenna McNally}, {Mona~Leigh Guha}, {Matthew~Louis Mauriello}, {and} {Allison
  Druin}. 2016.
\newblock \showarticletitle{Children's Perspectives on Ethical Issues
  Surrounding Their Past Involvement on a Participatory Design Team}. In {\em
  Proceedings of the 2016 CHI Conference on Human Factors in Computing
  Systems}. ACM, 3595--3606.
\newblock


\bibitem{DBLP:journals/tit/Miller56}
{George~A. Miller}. 1956.
\newblock \showarticletitle{Human memory and the storage of information}.
\newblock {\em {IRE} Trans. Information Theory\/} {2}, 3 (1956), 129--137.
\newblock
\showDOI{%
\url{http://dx.doi.org/10.1109/TIT.1956.1056815}}


\bibitem{mueller2011improving}
{Shane~T Mueller} {and} {Gary Klein}. 2011.
\newblock \showarticletitle{Improving users' mental models of intelligent
  software tools}.
\newblock {\em IEEE Intelligent Systems\/} {26}, 2 (2011), 77--83.
\newblock


\bibitem{nagel1998}
{{D.C.} Nagel}. 1998.
\newblock \showarticletitle{Human error in aviation Operations}.
\newblock {\em Human factors in Aviation, E.L.Weiner and E.C.Nagel (Eds)\/}
  19890047069, 34 (1998), 263--303.
\newblock
\showDOI{%
\url{http://dx.doi.org/10.1109/2.910904}}


\bibitem{DBLP:conf/chi/NielsenM90}
{Jakob Nielsen} {and} {Rolf Molich}. 1990.
\newblock \showarticletitle{Heuristic evaluation of user interfaces}. In {\em
  Proceedings of the Conference on Human Factors in Computing Systems, {CHI}
  1990}. 249--256.
\newblock
\showDOI{%
\url{http://dx.doi.org/10.1145/97243.97281}}


\bibitem{Norman:2002:DET:2187809}
{Donald~A. Norman}. 2002.
\newblock {\em The Design of Everyday Things}.
\newblock Basic Books, Inc., New York, NY, USA.
\newblock
\showISBNx{9780465067107}


\bibitem{SAEMS}
{Nat{\'a}lia Oliveira}, {Fabio Jorge}, {Jessica De~Souza}, {Valdir Junior},
  {and} {Leonardo Botega}. 2016.
\newblock {\em Development of a User Interface for the Enrichment of
  Situational Awareness in Emergency Management Systems}. Vol. 491.
\newblock 173--184.
\newblock
\showISBNx{978-3-319-41928-2}
\showDOI{%
\url{http://dx.doi.org/10.1007/978-3-319-41929-9_17}}


\bibitem{SA-Mines}
{E Onal}, {C Craddock}, {and} {Mica Endsley}. 2013.
\newblock \showarticletitle{From Theory to Practice: How Designing for
  Situation Awareness Can Transform Confusing, Overloaded Shovel Operator
  Interfaces, Reduce Costs, and Increase Safety}.
\newblock
\showDOI{%
\url{http://dx.doi.org/10.22260/ISARC2013/0171}}


\bibitem{PARUSH2017154}
{A. Parush}, {G. Mastoras}, {A. Bhandari}, {K. Momtahan}, {K. Day}, {B.
  Weitzman}, {B. Sohmer}, {A. Cwinn}, {S.J. Hamstra}, {and} {L. Calder"}. 2017.
\newblock \showarticletitle{Can teamwork and situational awareness (SA) in ED
  resuscitations be improved with a technological cognitive aid? Design and a
  pilot study of a team situation display}.
\newblock {\em Journal of Biomedical Informatics\/}  {76} (2017), 154 -- 161.
\newblock
\showISSN{1532-0464}
\showDOI{%
\url{http://dx.doi.org/10.1016/j.jbi.2017.10.009}}


\bibitem{DBLP:conf/ACMdis/Potts95}
{Colin Potts}. 1995.
\newblock \showarticletitle{Using Schematic Scenarios to Understand User
  Needs}. In {\em Proceedings of the 1st Conference on Designing Interactive
  Systems: Processes, Practices, Methods and Techniques, {DIS} '95}. 247--256.
\newblock
\showDOI{%
\url{http://dx.doi.org/10.1145/225434.225462}}


\bibitem{prasanna09}
{Raj Prasanna}, {Lili Yang}, {and} {Mlcolm King}. 2009.
\newblock \showarticletitle{Situation awareness oriented user interface design
  for fire emergency response}.
\newblock {\em Journal of Emergency Management\/}  {7} (03 2009), 65--74.
\newblock


\bibitem{DBLP:journals/corr/abs-1804-02767}
{Joseph Redmon} {and} {Ali Farhadi}. 2018.
\newblock \showarticletitle{YOLOv3: An Incremental Improvement}.
\newblock {\em CoRR\/}  {abs/1804.02767} (2018).
\newblock
\showURL{%
\url{http://arxiv.org/abs/1804.02767}}


\bibitem{DBLP:journals/thms/RegisDRTPCT14}
{Nicolas Regis}, {Fr{\'{e}}d{\'{e}}ric Dehais}, {Emmanuel Rachelson}, {Charles
  Thooris}, {Sergio Pizziol}, {Micka{\"{e}}l Causse}, {and} {Catherine
  Tessier}. 2014.
\newblock \showarticletitle{Formal Detection of Attentional Tunneling in Human
  Operator-Automation Interactions}.
\newblock {\em {IEEE} Trans. Human-Machine Systems\/} {44}, 3 (2014), 326--336.
\newblock
\showDOI{%
\url{http://dx.doi.org/10.1109/THMS.2014.2307258}}


\bibitem{DBLP:journals/ijmms/Richards00}
{Debbie Richards}. 2000.
\newblock \showarticletitle{The reuse of knowledge: a user-centred approach}.
\newblock {\em Int. J. Hum.-Comput. Stud.\/} {52}, 3 (2000), 553--579.
\newblock
\showDOI{%
\url{http://dx.doi.org/10.1006/ijhc.1999.0342}}


\bibitem{Webber73}
{{Horst W. J.} Rittel} {and} {{Melvin M.} Webber}. 1973.
\newblock \showarticletitle{Dilemmas in a general theory of planning}. In {\em
  Policy Sciences}, Vol.~4. 155--169.
\newblock


\bibitem{rushby2001modeling}
{John Rushby}. 2001.
\newblock \showarticletitle{Modeling the human in human factors}. In {\em
  International Conference on Computer Safety, Reliability, and Security}.
  Springer, 86--91.
\newblock


\bibitem{10.3389/frobt.2018.00015}
{Filippo Santoni~de Sio} {and} {Jeroen van~den Hoven}. 2018.
\newblock \showarticletitle{Meaningful Human Control over Autonomous Systems: A
  Philosophical Account}.
\newblock {\em Frontiers in Robotics and AI\/}  {5} (2018), 15.
\newblock
\showISSN{2296-9144}
\showDOI{%
\url{http://dx.doi.org/10.3389/frobt.2018.00015}}


\bibitem{Schuler:1993:PDP:563076}
{Douglas Schuler} {and} {Aki Namioka} (Eds.). 1993.
\newblock {\em Participatory Design: Principles and Practices}.
\newblock L. Erlbaum Associates Inc., Hillsdale, NJ, USA.
\newblock
\showISBNx{0805809511}


\bibitem{cps-survey}
{J. {Shi}}, {J. {Wan}}, {H. {Yan}}, {and} {H. {Suo}}. 2011.
\newblock \showarticletitle{A survey of Cyber-Physical Systems}. In {\em 2011
  International Conference on Wireless Communications and Signal Processing
  (WCSP)}. 1--6.
\newblock
\showDOI{%
\url{http://dx.doi.org/10.1109/WCSP.2011.6096958}}


\bibitem{demonhunt}
{Tim~Claudius Stratmann} {and} {Susanne Boll}. 2016.
\newblock \showarticletitle{Demon Hunt - The Role of Endsley's Demons of
  Situation Awareness in Maritime Accidents}. In {\em Human-Centered and
  Error-Resilient Systems Development}, {Cristian Bogdan}, {Jan Gulliksen},
  {Stefan Sauer}, {Peter Forbrig}, {Marco Winckler}, {Chris Johnson}, {Philippe
  Palanque}, {Regina Bernhaupt}, {and} {Filip Kis} (Eds.). Springer
  International Publishing, Cham, 203--212.
\newblock
\showISBNx{978-3-319-44902-9}


\bibitem{Tatar:2007:DTF:1466607.1466609}
{Deborah Tatar}. 2007.
\newblock \showarticletitle{The Design Tensions Framework}.
\newblock {\em Hum.-Comput. Interact.\/} {22}, 4 (Nov. 2007), 413--451.
\newblock
\showISSN{0737-0024}
\showURL{%
\url{http://dl.acm.org/citation.cfm?id=1466607.1466609}}


\bibitem{thomas2006general}
{David~R Thomas}. 2006.
\newblock \showarticletitle{A general inductive approach for analyzing
  qualitative evaluation data}.
\newblock {\em American journal of evaluation\/} {27}, 2 (2006), 237--246.
\newblock


\bibitem{usecaseSLR}
{Saurabh Tiwari} {and} {Atul Gupta}. 2015.
\newblock \showarticletitle{A Systematic Literature Review of Use Case
  Specifications Research}.
\newblock {\em Inf. Softw. Technol.\/} {67}, C (Nov. 2015), 128--158.
\newblock
\showISSN{0950-5849}
\showDOI{%
\url{http://dx.doi.org/10.1016/j.infsof.2015.06.004}}


\bibitem{DBLP:journals/tse/LamsweerdeDL98}
{Axel van Lamsweerde}, {Robert Darimont}, {and} {Emmanuel Letier}. 1998.
\newblock \showarticletitle{Managing Conflicts in Goal-Driven Requirements
  Engineering}.
\newblock {\em {IEEE} Trans. Software Eng.\/} {24}, 11 (1998), 908--926.
\newblock
\showDOI{%
\url{http://dx.doi.org/10.1109/32.730542}}


\bibitem{doi:10.1080/14639220701635470}
{Guy~H. Walker}, {Neville~A. Stanton}, {Paul~M. Salmon}, {and} {Daniel~P.
  Jenkins}. 2008.
\newblock \showarticletitle{A review of sociotechnical systems theory: a
  classic concept for new command and control paradigms}.
\newblock {\em Theoretical Issues in Ergonomics Science\/} {9}, 6 (2008),
  479--499.
\newblock
\showDOI{%
\url{http://dx.doi.org/10.1080/14639220701635470}}


\bibitem{attentionalwickens2009}
{Christopher~D Wickens} {and} {Amy~L Alexander}. 2009.
\newblock \showarticletitle{Attentional tunneling and task management in
  synthetic vision displays}.
\newblock {\em The International Journal of Aviation Psychology\/} {19}, 2
  (2009), 182--199.
\newblock


\end{thebibliography}
